\documentclass[aps,pra,showpacs,]{revtex4}
\begin{document}
\draft
\def\ds{\displaystyle}
\title{Inverted Oscillator}
\author{C. Yuce }
\address{ Physics Department, Anadolu University,
 Eskisehir, Turkey}
\email{cyuce@anadolu.edu.tr}
\author{A. Kilic }
\address{ Physics Department, Anadolu University,
 Eskisehir, Turkey}
\email{abkilic@anadolu.edu.tr}
\author{A. Coruh }
\address{ Physics Department, Sakarya University,
 Sakarya, Turkey}
\date{\today}
\begin{abstract}
The inverted harmonic oscillator problem is investigated quantum
mechanically. The exact wave function for the confined inverted
oscillator is obtained and it is shown that the associated energy
eigenvalues are discrete and it is given as a linear function of
the quantum number $n$.
\end{abstract}
\maketitle
\section{Introduction}

The exactly solvable potentials in quantum physics are very
limited. The harmonic oscillator problem is one of the rare
example of exactly solvable systems. Besides the mathematical
interest, the non-perturbative exact treatment plays a crucial
role in understanding and constructing for a new theory. For
example, the harmonic oscillator problem is used in high energy
physics and the state with zero-point energy is reinterpreted as
the vacuum.\\
If the frequency $\ds{\omega}$ is replaced with $\ds{i\omega}$,
the harmonic oscillator Hamiltonian becomes
\begin{equation}\label{hamilt}
H=-\frac{\partial^2 }{\partial x^2}-\omega^2 x^2~,
\end{equation}
where the constants are set to unity for simplicity
$\ds{(\frac{\hbar^2}{2m}=\hbar=1)}$. The potential in the new
hamiltonian is known as the inverted harmonic oscillator potential
or parabolic potential barrier. The inverted oscillator with an
exponentially increasing mass is known as Caldriola-Kanai
oscillator \cite{basco}. The inverted oscillator is the simplest
system whose solutions to Newton equations diverge exponentially
in phase space, a characteristic of chaotic motion.\\
Note that the replacement $\ds{\omega \rightarrow i\omega}$ can
not be applied to find the energy eigenvalues. If it can be made,
the energy eigenvalues would take complex values for the Hermitian
Hamiltonian (1). The reason not to apply it for finding the energy
spectrum is due to the  new boundary condition, namely the wave
function doesn't vanish at infinity under such a replacement.

The inverted harmonic oscillator problem (1) attracts great
attention not only for being one of the exactly solvable potential
in quantum mechanics \cite{p1,o1,o2,o3,o4,o5,o6,o7} but also
having wide range of application in many branches of physics. It
receives a record number of applications in many branches of
physics varying from high energy physics to solid state theory.
For example, it is remarkable that 2-d string theory can be mapped
on to the problem of non-interacting fermions in the inverted
harmonic potential. All the physics of 2-d string theory can be
recovered from this fermion theory. The massless tachyon field of
string theory is related to small fluctuations around the Fermi
surface \cite{string2}. Furthermore, the quantum sates of D0-brane
decay are precisely the quantum states of the Hamiltonian (1)
except that the spectrum starts at Fermi level and the model
provides us with a complete description of the dynamics of a
single D0-brane decay \cite{brane}. \\
The using of inverted harmonic oscillator in the context of
inflationary models was addressed by Guth and Pi \cite{guth}. In
that work, the authors used the inverted harmonic oscillator as a
toy model to describe the early time evolution of the inflation,
starting from a Gaussian quantum state centered on the maximum of
the potential.\\
Additionally, the parabolic potential with negative coupling (1)
is used to understand the thermal activation problem. The analogy
between the case of thermal activation and the case of the
one-loop effective potential for theories with spontaneous
symmetry breaking at tree level involve gaussian approximations
around unstable configurations \cite{thermal}. It was also shown
in \cite{thermal} that the problem of analyzing the decay of a
metastable state becomes effectively that of a quantum mechanical
inverted oscillator.\\
The inverted harmonic oscillator problem is also used as a model
of instability \cite{inst,inst2}. In the study of chaotic system,
$\omega^2$ is the instability parameter and determines the
unstable and stable directions the rate at which initial phase
space distributions expand and contract in these direction,
respectively \cite{inst}.\\
Another application of equation (1) is the statistical
fluctuations of fission dynamics which is studied by means of the
inverted oscillator \cite{fiss}.

The inverted harmonic oscillator is exactly solvable like standard
harmonic oscillator which plays an important role in constructing
the modern theories of physics. However, the inverted oscillator
has a continuous energy spectrum and there is no zero-point energy
associated with the inverted oscillator. It was also shown by many
authors that the energy eigenstates are no longer square
integrable.

In this study, we show that there is a direct link between the
free particle and inverted harmonic oscillator problems. We solve
the equation (1) analytically and it is shown that the inverted
oscillator admits discrete energy spectrum. Furthermore, it is
also shown that the wave function with discrete energy levels is
square integrable but there is no zero-point energy contrary to
standard harmonic oscillator.

\section{Formalism}

Let us now begin our study by introducing the following
transformation for the wave function
\begin{equation}\label{trans154}
\Psi(x,t)=\exp \left( \frac{i\omega x^2}{2} -\omega t \right)
\Phi(x,t)~.
\end{equation}
By substituting the equation (\ref{trans154}) into the Schrodinger
equation with the Hamiltonian (1), we get the transformed
equation.
\begin{equation}\label{deriv}
-\frac{\partial^2 \Phi}{\partial x^2}-2 i \omega x ~
\frac{\partial \Phi}{\partial x} =i
 \frac{\partial \Phi}{\partial t} ~.
\end{equation}
A canonical transformation is introduced as follows
\begin{equation}\label{cgh}
x= q(t)~y~,
\end{equation}
where $q(t)$ satisfies the classical equation of motion for the
inverted harmonic oscillator potential. By solving the
corresponding Lagrange equation with the potential $\ds{-\omega^2
x^2}$, the equation of motion is obtained
\begin{equation}\label{classic}
\frac{\ddot{q}(t)}{2}=2\omega^2 q(t)~.
\end{equation}
Since the mass was set $\ds{(2m=1)}$, $1/2$ factor is introduced
in front of $\ds{\ddot{q}}$ instead of writing mass directly.
Then, the solution follows
\begin{equation}
q(t)=e^{ 2 \omega t}~.
\end{equation}
The canonical transformation (\ref{cgh}) rescales the coordinate.
The scaling parameter is given by solution of the corresponding
Lagrange equation. It is just the classical path.\\
Under the transformation (\ref{cgh}), the time derivative operator
transforms as
\begin{equation}\label{jef}
\frac{\partial}{\partial t} \rightarrow \frac{\partial}{\partial
t}-2 \omega y \frac{\partial}{\partial y}~.
\end{equation}
If we substitute the equations (\ref{cgh},\ref{jef}) into the
equation  (\ref{deriv}), we obtain
\begin{equation}
-\frac{\partial^2 \Phi}{\partial y^2}=i e^{4 \omega t}~
\frac{\partial \Phi}{\partial t} ~.
\end{equation}
We can rewrite the above equation just by using the separation of
variables technique
\begin{equation}\label{parametr}
\Phi(y,t)=\exp {(~- \frac{i\epsilon}{4 \omega} e^{-4 \omega t}~) }
\Phi(y)~.
\end{equation}
where $\ds{\epsilon}$ is a constant and $\ds{\Phi (y)}$ is given
by
\begin{equation}
-\frac{\partial^2 \Phi (y)}{\partial y^2}= -\epsilon ~\Phi (y)~ .
\end{equation}
It is very interesting to observe that the new equation is just
the Schrodinger equation for the free particle. A canonical
transformation (\ref{cgh}) connects the two Hamiltonians. So we
conclude that the dynamical properties of the inverted harmonic
oscillator and free particle can be inferred from each other.\\
By using the well-known wave functions for the free particle, we
can obtain the wave functions for the inverted harmonic
oscillator. There are two solutions of the free particle: the
plane wave solution and the solution for which the particle is
confined into a box. So, the two solutions for the inverted
harmonic oscillator can be constructed as follows by using the
equations (2,\ref{cgh},\ref{parametr})
\begin{equation}\label{sol1}
\Psi_1(x,t)=N \exp \left( \frac{i\omega x^2}{2} -\omega t ~-
\frac{i\epsilon}{4 \omega} e^{-4 \omega t}~\right)  \sin{(e^{-2
\omega t} \sqrt{\epsilon} ~x)}~,
\end{equation}
\begin{equation}\label{sol22}
\Psi_2(x,t)=\exp \left( \frac{i\omega x^2}{2}+ik~e^{-2 \omega t}
~x -\omega t ~+ \frac{i k^2}{4 \omega} e^{-4 \omega t} ~\right)~,
\end{equation}
where $N$ is a normalization constant and a new constant $k$ for
$\ds{\Psi_2}$ is given by $\ds{k^2=-\epsilon}$ . There is no
normalization constant for $\ds{\Psi_2}$, since the plane wave
solution from
which  $\ds{\Psi_2}$ is constructed can not be normalized.\\
The Hamiltonian (1) is invariant under the parity operator. It can
be seen that $\ds{\Psi_1}$ is an eigenket of the parity operator.
It has an odd-parity. This is not the case for the standard
harmonic oscillator, since the hermite polynomials $\ds{H_n}$ have
the odd-parity for the
odd-numbers of $n$ and the even-parity for the even-numbers of $n$.\\
One another interesting case is the application of the time
reversal operator $\ds{(t\rightarrow -t, i\rightarrow -i)}$. We
see from the equations (\ref{sol1},\ref{sol22}) that time reversal
operation is equivalent to the following replacement:
$\ds{\omega\rightarrow - \omega}$. Fortunately, the Hamiltonian
(1) is invariant under such a replacement. So, the wave functions
under the application of the time reversal operator are also the
solution.\\
Now, let us interpret the first solution (\ref{sol1}) from the
physical point of view. The sinusoidal character of $\ds{\Psi_1}$
plays an important role since the confinement of the inverted
oscillator can be achieved by this function. The particle can be
confined in an expanding box. The stationary boundary condition is
transformed to the moving boundary condition by the equation
(\ref{cgh}). The wave function is zero at both the origin and
$\ds{x=L_0 e^{2\omega t}}$, where $\ds{L_0}$ is the initial
length. So, the constant $\epsilon$ in (\ref{sol1}) is given by
($\ds{\sqrt{\epsilon}=\frac{n \pi}{L_0}}$). Note that this
constant doesn't coincide the energy eigenvalues exactly, as can be seen below.\\
An important observation is made for the time-dependent function
$\ds{q(t)=e^{2\omega t}}$. It is just the particle position for
the inverted oscillator in classical mechanics. However, in
quantum mechanics, it is the position of the wall surrounding the
particle. In other words, it is the boundary condition. The speed
of the particle in classic physics and the the speed of the wall
are exactly the same. In classical mechanics, the particle is
exactly at the position $\ds{q(t)=e^{2\omega t}}$, however in
quantum mechanics, it may be found in the interval between the
origin and the position $\ds{q(t)=e^{2\omega t }}$ .\\
Since the wall is expanding, one may say that the transition
between the states may occur during the expansion. But, this is
not the case, since the wave functions $\ds{\Psi_1}$ for different
values of $n$ are orthogonal to each other. No transition occurs
during the expansion of the
wall.\\
Now, let us normalize the wave function  $\ds{\Psi_1}$ and then
find it's the energy eigenvalues.
\begin{equation}
\int_0^{(L_0 e^{2 \omega t})} |\Psi_1|^2 dx=|N|^2 \frac{L}{2}=1~.
\end{equation}
It is interesting to observe that the the normalized wave function
can be found for the inverted harmonic oscillator in contrary to
the existing idea. However, the second solution (\ref{sol22}) is
not normalized. The second solution is found to decay with the
decay rate $\ds{4\omega}$
\begin{equation}
 |\Psi_2|^2=e^{-2\omega t}=e^{-\frac{\Gamma}{2} t}~,
\end{equation}
where $\ds{\Gamma}$ is the decay rate. Having obtained the
normalization constant, we can now compute the energy eigenvalues
for $\ds{\Psi_1}$. It is given by
\begin{equation}\label{energy}
E_n=\int_0^{(L_0 e^{2 \omega t})} \Psi_1 H \Psi_1 dx=e^{-4 \omega
t} \frac{n^2 \pi^2}{L_0^2}~.
\end{equation}
The energy eigenvalues are decreasing as time goes on. The
inverted oscillator admits discrete energy levels and it has a
strong resemblance to free particle's energy spectrum. At the
initial time, the energy eigenvalues for the inverted oscillator
coincide with those of the free particle.As a special case, if
$\ds{\omega}$ is set to zero, then energy eigenvalues are reduced
to those of the
free particle as it is expected.\\
There is no zero energy associated with the inverted oscillator in
contrary to the standard harmonic oscillator. The energy depends
on the quantum number linearly for the standard oscillator and
quadratically for the confined inverted oscillator. \\
As an application, consider a quantum gas contained in an inverted
harmonic oscillator potential. Let the quantum gas be confined in
a box. One can investigate the problem statistically. Now, assume
that the length (volume in 3D) of the box is increased
exponentially. Then, the energy of each individual atom is
decreased (\ref{energy}). During the expansion, no transition
between the states in the statistical system occurs as it was
explained before. Since the temperature of the system is related
to the energy, we can safely say that the temperature is decreased
when the volume is increased. The quantum gas gets cooler when the
box expands. The dependence of temperature on volume for the
inverted oscillator can be derived by using the relation
(\ref{energy}) and the laws of statistical physics. The model
studied here may be used for the non-equilibrium statistical
physics.

\end{document}